\documentclass[prl,twocolumn]{revtex4}
\usepackage{epsfig,amssymb,amsmath,latexsym,color,pifont,graphicx,psfrag}
\newcommand{\p}{\varphi}
\newcommand{\e}{\varepsilon}
\newcommand{\pp}{\boldsymbol\p}
\newcommand{\om}{\omega}
\newcommand{\er}{\eqref} 
\DeclareMathOperator{\sgn}{sgn}

\begin{document}

\title{Network reconstruction from random phase-resetting}

\begin{abstract}
We propose a novel method of reconstructing the topology and interaction functions for a general oscillator network. An ensemble of initial phases and the corresponding instantaneous frequencies is constructed by repeating random phase-resets of the system dynamics. The desired details of network structure are then revealed by appropriately averaging over the ensemble. The method is applicable for a wide class of networks with arbitrary emergent dynamics, including full synchrony.
\end{abstract}

\author{Zoran Levnaji\'c}
\author{Arkady Pikovsky}
\affiliation{Department of Physics and Astronomy, University of Potsdam, 14476 Potsdam, Germany}
\maketitle

Complex networks of many interacting units found at all scales in nature are the subject of intense research in many scientific areas~\cite{boccal-costa}. Among the central issues in this field are the exploration and development of methods for determining the architecture of a network based on the observable data. Knowing the network structure helps in understanding its collective behavior, and indicates ways to engineer networks with the desired properties. For instance, it has been realized that inferring the topology of gene regulatory networks is crucial for completing our knowledge about the inner workings of cells~\cite{barabasi-ja}. Many real networks display modular and community structure that is essential for their functioning~\cite{milo} and can be extracted using a variety of methods~\cite{fortunato}, as done for yeast metabolic network~\cite{herrgard}. Reconstruction techniques often rely on examining the time-series of network dynamics that can reveal its interaction functions~\cite{tkr}. The network topology can be detected by studying the interchanges among its collective behaviors or investigating its response dynamics~\cite{timme}. Structural properties can be determined from various time-scales in the emergence of synchronization~\cite{arenas-prk}, or by employing specific control theory methods~\cite{dongchuan}. Recently proposed techniques involve noisy dynamical correlations between the nodes~\cite{ren}, and even tackle models with non-equilibrium dynamics~\cite{roudi}.

However, existing reconstruction methods, that often use network models with single-node dynamics represented by different types of oscillators~\cite{kuramoto-acebron}, typically require long time-series of dynamical data, or a certain level of complexity in the emergent dynamics~\cite{tkr,timme}. Since synchronization destroys the initial node-related information, detecting network topology in such cases is extremely difficult. Some methods are applicable only to sparse or non-directed networks, often providing results with only a limited precision~\cite{ren}.

In this Letter, we propose a novel method of reconstructing the topology and interactions of a general oscillator network. Our idea relies on repeatedly re-initializing the network dynamics (e.g. by performing random \textit{phase-resets}), in order to produce an ensemble of the initial dynamical data. We design the quantities obtained by averaging this ensemble, whose values reveal the desired details of the network structure. Our method is applicable to any directed and weighted network, with general interaction functions and oscillator frequencies, and with arbitrary emergent dynamics, while avoiding the need for long time-series.

In the context of phase-resets, one is typically interested in the \textit{phase-resetting curves}, which specify the system's response to weak external perturbations~\cite{preyer-tass}. They have been investigated both experimentally~\cite{galan-ramon} and theoretically~\cite{ko-kori,us}, and been shown to contain properties relevant for determining network details such as clustering~\cite{achuthan}. An algorithm for the estimation of neuron interaction and its stability based on phase-resets has been proposed~\cite{galan-ramon}. We here employ phase-resetting somewhat differently, since our interest lies in the \textit{internal} network interactions, rather than its response to stimuli. Contrary to~\cite{galan-ramon}, we use phase-resets only as a natural way to re-initialize the dynamics of an oscillator network, without measuring the phase shifts occurring due to resetting.

Our model consists of $N$ oscillators (nodes), characterized by their phases $\p_i \in [0,2\pi)$ and natural frequencies $\om_i$. They are coupled pair-wise, via general $2\pi$-periodic interaction functions $f_{ij}$ with zero mean:
\begin{equation} \dot{\p}_i = \om_i + \sum_{j=1}^{N}  f_{ji} (\p_j - \p_i ) \; . \label{eq-1} \end{equation}
Models of this type include the famous Kuramoto model and its generalizations, widely used in theoretical studies, as well as for describing specific experimental situations~\cite{arenas-prk,kuramoto-acebron,galan-ramon}. The functions $f_{ij}(\phi)$ are generally non-symmetric with respect to exchange of indices, and thus fully define the dynamical network (order of indices determines the direction of interaction). Network adjacency matrix given as $A_{ij}=\sgn |f_{ij}|$ specifies its topology. Dynamics starts from a set of initial phases (i.p.) which we denote as $\pp=(\p_1,\hdots \p_N)(t=0)$, chosen from a distribution $\rho(\pp) > 0$ normalized to $(2\pi)^N$. The method is based on two assumptions:
\textit{(i)} we are able to arbitrarily re-initialize the network dynamics $I$ times, by independently resetting the phases of all nodes to a new state $\pp$;  
\textit{(ii)} we are able to measure all the values $\pp_l$, and all initial instantaneous frequencies $\dot\pp_l$, each time the dynamics is re-initialized (for $l=1,\hdots I$).
As we show in what follows, the ensemble of data for $I \gg 1$ created under these assumptions yields the entire network structure.

Introducing a $2\pi$-periodic \textit{test-function} $g=g(\p_i - \p_j)$ with zero mean, our aim is to compute the \textit{reconstruction index} $S_{ij}$ defined as:
\begin{equation}  S_{ij} [g] =(2\pi)^{-N} \int_{[0,2\pi]^N}   d\pp \; g(\p_i - \p_j) {\dot\p}_j(\pp)   \; .  \label{eq-defS} \end{equation}
Taking the functions $f_{ij}$ in Eq.\er{eq-1} to be generally given by the Fourier series $f_{ij}(\phi) = \sum_n a^{(n)}_{ij} \sin n \phi + b^{(n)}_{ij} \cos n \phi$, we obtain the following expression for $S_{ij}$:
\[ \begin{gathered}
S_{ij}  =  (2\pi)^{-N} \; \sum_{k=1}^{N} \sum_{n=1}^{\infty}  \int_{[0,2\pi]^N}d\pp \; g(\p_i - \p_j) \; \times \;\;\;\;\;  \\
\;\;\;\;\;\;\;\;\;\;\;\;   \big[ a^{(n)}_{kj} \sin (n\p_k - n\p_j) +   b^{(n)}_{kj}  \cos (n\p_k - n\p_j) \big] \; ,
\end{gathered} \] 
which is independent of the frequencies $\om_i$. The integral over $\p_i$ vanishes unless $i=k$. This implies that if $A_{ij}=0$, the corresponding $S_{ij}=0$, independently of the choice of $g$. The non-zero entries of $S_{ij}$ directly reveal the presence of network links. In addition, matrix $S_{ij}$ detects the desired properties of the interaction functions for appropriately selected test-function $g$. In particular, using $g(\phi)=2 e^{i n \phi}$ we obtain the Fourier harmonics of $f_{ij}$, which are the interaction parameters $a^{(n)}_{ij}$ and $b^{(n)}_{ij}$:
\begin{equation} S_{ij} [2 e^{i n \phi}] = b^{(n)}_{ij} + i a^{(n)}_{ij}= \frac{1}{\pi} \int_0^{2\pi}f_{ij}(\phi)e^{i n \phi} d\phi \; . \label{eq-sincos} \end{equation}
Computation of $S_{ij}$ for adequate $g$ amounts for reconstruction of any dynamical network described by Eq.\er{eq-1}. Depending on the properties of $f_{ij}$ that are to be examined, other choices of $g$ are also possible. When dealing with the empirical interaction functions involving an unknown number of Fourier harmonics, a specifically designed $g$ based on the experimental assumptions about $f_{ij}$ might be useful.
This result is largely independent of the frequencies $\om_i$, the network's directedness, and the distribution $\rho$. In particular, it is also independent of the network's final dynamical state, whether dependent on $\rho$ or not. However, a constant component in case of $f_{ij}$ with non-zero mean cannot be detected, since its presence is indistinguishable from the natural frequency $\om$.

To practically implement our method, we need to convert the integral from Eq.\er{eq-defS} into an average involving discrete non-uniformly distributed empirical data $\{ \pp_l \}_{l=1}^{I}$ and $\{ \dot\pp_l \}_{l=1}^{I}$. To that end, we represent the function $\dot{\p}_j(\pp)$ using the kernel smoother $Q(\pp-\pp_l)$~\cite{wand-scott} as:
\[ \dot{\p}_j(\pp) = \frac{\sum_{l=1}^I Q(\pp-\pp_l) \dot{\p}_j(\pp_l)}{\sum_{l=1}^I Q(\pp-\pp_l)} \; . \]
The denominator is just the empirical density $\rho(\pp) = \sum_l Q(\pp-\pp_l)$ obtained via kernel distribution estimate~\cite{wand-scott}. Since the integration over $\pp$ already provides smoothing, we take $Q(\pp-\pp_l)\to\delta(\pp-\pp_l)$, and replace the Eq.\er{eq-defS} with a practical formula for $S_{ij}$:
\begin{equation} S_{ij} [g] = \bigg\langle \dfrac{{\dot\p}_j  g(\p_i - \p_j)}{\rho(\pp)} \bigg\rangle = \frac{1}{I} \sum_{l=1}^I \frac{\dot{\p}_j(\pp_l) g(\p_i - \p_j)}{\rho(\pp_l)} \; , \label{eq-practicalS} \end{equation}
which is the average of empirical ${\dot\p}_j g$ weighted by $\frac{1}{\rho}$.

The most trivial way to obtain the ensemble $\{ \pp_l \}_{l=1}^{I}$ would be to pick the values from a fixed distribution $\rho(\pp)$. Instead, we seek to mimic an experimentally feasible situation by performing $I$ random phase-resets of the network dynamics, separated by the time interval $\tau$.  Mathematically, this amounts to adding the term $\sum_{l=1}^{I} K_{i,l} \sin (\p_i + \alpha_{i,l}) \delta (t - l\tau)$ to the RHS of Eq.\er{eq-1}~\cite{us}. For each reset $l$ and each oscillator $i$, we independently pick the kicking strength $K_{i,l}$ from a zero mean Gaussian distribution with standard deviation $K=1$, and the phase-shift $\alpha_{i,l}$ uniformly from $[0,2\pi)$. The ensemble is constructed by storing the phase values immediately after resets. The resulting artificially created ensemble has little in common with the natural distribution of phases, and can be considered as approximately independent. This is expressed by separability of $\rho(\pp)$ into a product of $N$ one-dimensional distributions $\rho_i (\p_i)$:
\begin{equation} \rho(\pp) = \prod_{i=1}^N  \rho_i (\p_i) \; , \label{eq-dd}  \end{equation} 
each of which we determine from generated data using the kernel estimation method~\cite{ourkernel}. After each reset, the ensemble of $\dot\pp$ is computed using a small time interval. The phase value prior to reset is of no importance, since our interest is not in the phase-resetting curves, but in modeling a realistic way to create the ensemble $\pp$.
The described procedure is quite similar to the recent experimental implementation of the randomized phase-resetting of epileptic neurons aimed at their transient desynchronization~\cite{hauptmann-tass}. In these experiments, however, the problem of simultaneous read-out of phases and frequencies remains a challenge.

\begin{figure}[!ht] \centering
\includegraphics[width=0.775\columnwidth]{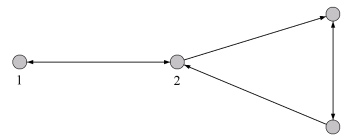}   \caption{4-node network used for illustrating our method.}
\label{fig-1} \end{figure}

\begin{figure}[!ht] \centering
\includegraphics[width=0.885\columnwidth]{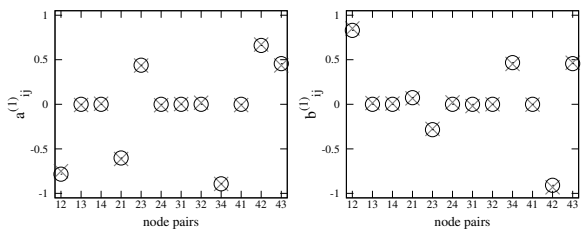} \caption{Reconstruction of the network from Fig.\ref{fig-1}. Circles: actual parameter values, crosses: numerically obtained values for $I=10^4$. 
Left: $a^{(1)}_{ij}$, right: $b^{(1)}_{ij}$, for each node pair $i\to j$.} 
\label{fig-2} \end{figure}

We now illustrate our theoretical findings through numerical simulations on simple network examples, computing the reconstruction index $S_{ij}$ as described above. Consider a simple network with $N=4$ oscillators shown in Fig.\ref{fig-1}. 
We pick the natural frequencies at random from $\om_i \in [-1,1]$. The interaction functions $f_{ij}$ are defined for linked node pairs by randomly choosing $a^{(1)}_{ij}, b^{(1)}_{ij} \in [-1,1]$, while taking $a^{(n)}_{ij}=b^{(n)}_{ij}=0$ for $n \ge 2$. Since such a network typically does not synchronize, our approximation of independent i.p. after resetting is appropriate. We take $g=2 e^{i\phi}$ and compute $S_{ij}$ from an ensemble of $I=10^4$ i.p. to obtain the numerical approximations of $a^{(1)}_{ij}$ and $b^{(1)}_{ij}$ via Eq.\er{eq-practicalS}. In Fig.\ref{fig-2} we compare the numerical $a^{(1)}_{ij}$ and $b^{(1)}_{ij}$ (crosses) with the actual values (circles).
All values display a very good agreement for both linked node pairs (different from zero) and non-linked node pairs (zero). We have not only revealed the adjacency matrix $A_{ij}$, but also found the interaction parameters $a^{(1)}_{ij}$ and $b^{(1)}_{ij}$, thus reconstructing the entire dynamical network.

\begin{figure}[!ht] \centering
\includegraphics[width=0.885\columnwidth]{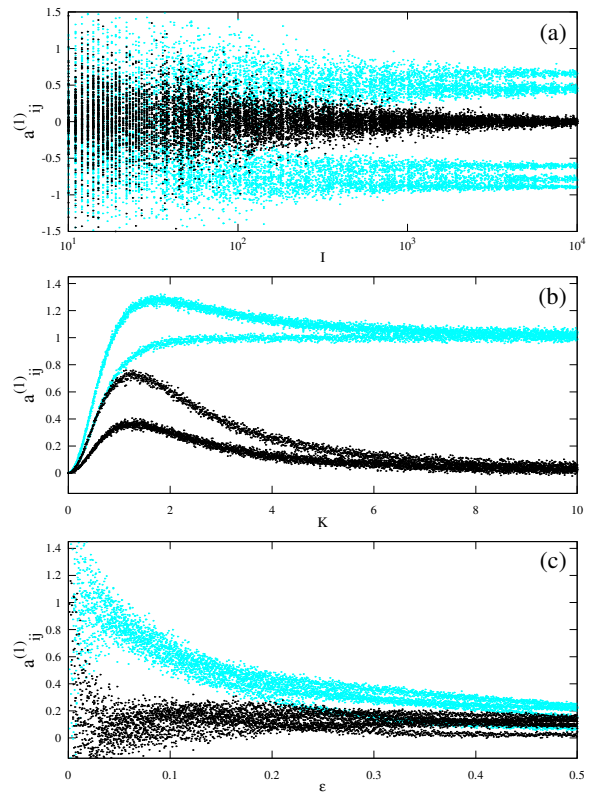} \caption{(color online). Numerical values of $a^{(1)}_{ij}$ for network in Fig.\ref{fig-1}, for links (cyan/gray) and non-links (black). (a) computed from ensemble of $I$ i.p. (cf. Fig.\ref{fig-2}). (b) computed from $I=10^4$ for network with attractive interactions, and with resetting done at synchronous state using kicking strength $K$. (c) computed from $I=10^5$ for network with attractive interactions where only spikes ($\p=0$) are observable, in relation to coupling strength $\e$ (see text for details).}
\label{fig-3} \end{figure}

Below we discuss the limitations of our method. If the available data ensemble $I$ is too small, the statistics is poor and the obtained network characteristics have large uncertainties, which typically decrease as $\sim I^{-\frac{1}{2}}$. To illustrate this, in Fig.\ref{fig-3}a we present the numerical values of parameter $a^{(1)}_{ij}$, computed for network in Fig.\ref{fig-1} using the ensemble of i.p. $\pp$ of size $I$. While the distinction between links and non-links can already be seen for $I \sim 10^3$, for good approximation one needs $I \gtrsim 10^4$ (as done in Fig.\ref{fig-2}). For higher Fourier harmonics, the convergence is gradually slower, but maintains the same properties.

Another limitation is related to the validity of our independence assumption for the ensemble of i.p. which is expressed by the separability of distribution $\rho (\pp)$ Eq.\er{eq-dd}. This heavily depends on the network's dynamical regime and the resetting strength. For a full synchrony and weak kicking, the reset state is expected to be strongly correlated, whereas for chaotic dynamics and strong resets, the independence assumption is essentially correct. To study this, we consider again the network from Fig.\ref{fig-1}, but now we fix all frequencies to $\om_i = 1$, and take all interactions to be attractive $a^{(1)}_{ij}=1, b^{(1)}_{ij}=0$ (Kuramoto-type model with identical oscillators). We apply random kicking as described above after allowing the network to synchronize ($\tau \gg t_{synch}$), but this time with a variable standard deviation of kicking strength $0 < K < 10$. For each value of $K$ we create an ensemble of $I=10^4$ i.p., and use it to compute $a^{(1)}_{ij}$ as done previously. In Fig.\ref{fig-3}b we show the reconstructed values of $a^{(1)}_{ij}$ for links and non-links in relation to $K$. Sufficiently strong kicking ($K \gtrsim 5$) succeeds in destroying the network's synchrony and generating the independent i.p., from which a good approximation of $a^{(1)}_{ij}$ is computed. Moderate kicking $K \sim 1$ applied previously are now insufficient. This furthermore depends on the relation between $\tau$ and $t_{synch}$: if $\tau \lesssim t_{synch}$ (frequent resets) the separability of $\rho$ is easier to achieve. Too strong kicking can also induce correlations in $\pp$, regardless of dynamical regime and $\tau$. However, note that $\rho$ can be estimated using the techniques more elaborate than simple one-dimensional kernels~\cite{wand-scott}, which can in principle yield a good estimate even in the non-separable case. On the other hand, phase-resetting is potentially not the only mechanism of obtaining the ensemble $\pp$; recall that our theory with a known $\rho(\pp)$ works equally well for any case, including full synchrony and inseparability.

Adding noise terms to RHS of Eq.\er{eq-1} does not formally change the derivation of our main result, rendering our theory valid in the presence of noise. However, in the light of discussion above, noise will have an effect on the performance of method: additional uncertainty due to larger fluctuations of the estimated $\dot\pp$ require larger ensembles to achieve the desired precision. On the other hand, noise may play a constructive role by destroying the undesired correlations within $\pp$, and thus facilitating the separability of $\rho$.

While the experimental techniques for measuring $\pp$ are already in use~\cite{galan-ramon}, in a potential realistic application of our method a problem may arise in relation to the measurement of $\dot\pp$. The entire cycle of a real oscillator is often not accessible; instead, one can observe only a single event per period (e.g. a spike produced by a neuron). In such cases, one is forced to estimate the instantaneous frequencies relying solely on the time intervals between the spikes. To illustrate this, we consider again the system studied in Fig.\ref{fig-3}b, but now we replace $a^{(1)}_{ij}$ with $\e a^{(1)}_{ij}$. The parameter $\e$ (coupling strength) controls the ratio between the oscillation time-scale (period) and the interaction time-scale (synchronization). Rather than computing instantaneous $\dot\pp$ after each reset, we observe only the event of an oscillator passing through the phase value $\p=0$ (spike), and estimate both $\pp$ and $\dot\pp$ from the first two spikes observed after resetting. We then reconstruct the values of $a^{(1)}_{ij}$ using the ensemble of $I=10^5$ i.p. as done before (strong resetting is applied immediately after the spikes are recorded). The results shown in Fig.\ref{fig-3}c have a clear physical interpretation: for too small coupling $\e \lesssim 0.03$ the links can not be revealed since the interaction is too weak. For too large $\e \gtrsim 0.4$ the two time-scales are too close, and the detection is again impossible since the distribution of phases changes significantly over a period. However, between these extremes, there is a range of coupling around $\e \sim 0.1$ where the two time-scales are well separated allowing a reliable reconstruction. This shows that with an adequately big ensemble our method works even if the entire oscillator cycle is not accessible: errors in the estimation of $\pp$ and $\dot\pp$ play a role similar to the noise. The method fails in the case of too strong coupling, similarly to the case of too weak resetting after synchronization (cf. Fig.\ref{fig-3}b).

In conclusion, we proposed a method of reconstructing oscillator networks by repeating random phase-resets, applicable to a general network irrespectively of the dynamical regimes (the feasibility of such resetting has been recently demonstrated for neural tissue~\cite{hauptmann-tass}). Our theory emphasizes the importance of the \textit{transient dynamics} in the context of network reconstruction, thus complementing the available techniques that rely on time-series recorded in final stationary state.
Our theoretical model can be straightforwardly generalized to other models beyond Eq.\er{eq-1}. If the couplings depend on two phases in a more general way, or depend on more than two phases, one should use more elaborate test-functions (e.g. in a form of general complex exponentials); however, even a theoretical description of such networks is already a challenge. For high-dimensional oscillators only a single scalar might be observable: our method can still be applied through the appropriate transformation to phases~\cite{tkr}. Another generalization regards the reconstruction of sub-networks, in the case that only information on some nodes is accessible. The problem here is to infer the distribution of i.p. for the non-accessible nodes. Finally, a real experimental situation may involve a network whose dynamics cannot be reset for all nodes simultaneously, which renders the independence assumption invalid. This is a much more challenging, although very realistic case that requires additional study.

\acknowledgments Support from DFG via project FOR868 is acknowledged. Thanks to A.~D\'iaz-Guilera for useful discussions.


\begin{thebibliography}{}
\bibitem{ourkernel} Using the kernel $Q = e^{\sigma (\p - \p_l)}$, we find for each node $i$ the optimal value $\sigma_i$ by cross-validation via maximal likelihood method~\cite{wand-scott}. This value defines 1-D  distribution estimate $\rho_i (\p_i) = \sum_l Q(\p_i - \p_l) = \sum_l  e^{\sigma_i (\p_i - \p_l)}$ for this node, based on the empirical reset ensemble $\{ \p_l \}_{l=1}^I$.
\end{thebibliography}
\end{document}